\documentclass[11pt,letter]{article}

\usepackage[utf8]{inputenc}
\usepackage[T1]{fontenc}
\usepackage[margin=1in]{geometry}
\usepackage{setspace}
\usepackage{amsmath,amssymb}
\usepackage{titlesec}
\usepackage[hidelinks]{hyperref}

\titleformat{\section}{\normalfont\large\bfseries}{\thesection.}{0.6em}{}
\titlespacing*{\section}{0pt}{1.2em}{0.6em}

\title{The NISQ Trap:\\
\large\itshape Eight Years of Demonstrations the Hardware Was Built to Lose}

\author{Amit Hagar\thanks{Department of History and Philosophy of
Science and Medicine, Indiana University, Bloomington, IN 47405.
Email: \texttt{hagara@iu.edu}. This paper extends the argument the author made in
\emph{The Complexity of Noise} (Springer Nature, 2010), consolidating
a five-part Substack series on NISQ de-quantization results; it was
drafted with AI assistance and substantially edited by the author,
and all claims and references were checked by hand.}}

\date{\today}

\begin{document}

\maketitle

\begin{abstract}
\noindent With a single clear exception, every NISQ-era flagship demonstration of ``quantum advantage'' has, within eighteen months of its announcement, been classically reproduced, shown to rest on classically tractable structure, or closed by a simulability theorem. Six theoretical results from 2024 through April 2026 explain the pattern: the regions of circuit-space NISQ hardware can run with sufficient fidelity coincide with the regions classical algorithms compress efficiently, because the features that admit one (low effective depth, strong algebraic structure, geometric locality) are the features that admit the other. Reconstructing the NISQ programme in this light, this article dates it from its 2018 articulation as an interim retreat from the unmet conditions of the 1996 threshold theorems, and characterises the eight years that followed as a closed loop in which the demonstrations the hardware could run were drawn from regions classical methods could already reach. The exit from the loop is located where the threshold theorems originally located it: in fault tolerance. The empirical pattern could in principle break with a demonstration that escapes the current simulability results. After eight years and more than thirty advantage-class announcements, the burden of producing such a demonstration falls to the defenders of NISQ.
\end{abstract}

\section{The two halves of one constraint}

In April 2026, four authors posted a paper on the arXiv giving a classical algorithm for the fermionic-dynamics demonstration recently run on a trapped-ion quantum computer~\cite{Alam2025,Oh2026}. The demonstration had been presented, in its title, as evidence of dynamics beyond exact classical simulation, and by the vendor as a processor that had ``departed the era of classical simulation.'' The classical algorithm reproduces the noninteracting sector of the experiment to additive error, including the high-weight Wilson-loop observables previously thought to lack any reliable classical reference; there it matches the exact benchmarks where they exist and sides with the tensor-network predictions against the error-mitigated experimental curves. It does not reach the interacting regime: interactions break the free-fermionic dynamics, the sampling cost grows prohibitive well before the experiment's parameter range, and the experiment's interacting endpoint would require on the order of $10^{25}$ samples by the authors' own estimate. The experiment's central claim, as stated in its title, therefore stands: it remains beyond exact classical simulation, and what the classical algorithm supplies is additive-error estimation, the same precision a finite-shot quantum device achieves. What the classical result exposed is narrower and more telling. The structured non-Gaussian input states the experiment used are tensor products of disjoint two-fermion pairs, mathematically identical to the canonical magic resource for fermionic linear optics. Such states admit an algebraic compression: an exponentially large sum of Slater determinants collapses into a coefficient of a multivariate Pfaffian polynomial, extractable by randomized sign-averaging. That same structure, chosen because the hardware could run it, is the structure the classical algorithm exploited to follow the experiment as far as the free dynamics preserved it. The boundary moved where the structure allowed, and stopped where the structure ran out.

This is the latest in a sequence. Six months earlier, Mele, Angrisani and collaborators proved that under standard local-noise models above a constant rate, any quantum circuit converges, in the statistics of any measurable observable, to a circuit of logarithmic effective depth~\cite{Mele2025}. The early layers of the computation are erased by the noise; only the final $O(\log n)$ layers leave a trace in the measurement statistics. Concurrent work by Nelson, Rajakumar and Gullans tightened the bound under geometric locality, the constraint every superconducting and trapped-ion architecture in fact satisfies~\cite{Nelson2025}. Independent work established classical simulability of noisy circuits in quasi-polynomial time under approximate Markovianity~\cite{Markov2025}, and analogous results for bosonic circuits showed that non-Gaussianity, the very resource credited with bosonic quantum advantage, accelerates classical attack once noise is present~\cite{Bosonic2025}. Earlier work by Oh on linear optics had already established percolation-based classical simulability of constant-depth circuits with photon loss~\cite{Oh2024}.

Each paper carves off a region of circuit-space and proves it classically tractable. Read individually, each is a technical result about a particular regime. Read together, they describe a single constraint with two faces. One face is hardware: noisy processors cannot maintain coherence across deep unstructured circuits and so must run shallow circuits or circuits with strong algebraic structure (paired states, Gaussian dynamics, low-weight observables, integrable Hamiltonians). The other face is algorithmic: shallow circuits and circuits with strong algebraic structure are exactly the circuits classical methods compress efficiently. The two faces meet at a sharp identity. The hardware that can run a circuit and the classical algorithm that can simulate it are responding to the same property of the circuit.

\section{The closed loop}

Call this the closed loop. A NISQ demonstration of quantum advantage requires three ingredients: a circuit the hardware can execute with sufficient fidelity, an observable the hardware can measure to useful precision, and a classical baseline the result outperforms. The hardware constraint forces the first two ingredients into the structured/shallow region; the simulability theorems show that the structured/shallow region is the region within which the third ingredient cannot be supplied. The loop is a direct consequence of three independently established facts. Noisy hardware caps achievable circuit complexity at logarithmic effective depth~\cite{Mele2025}; structured initial states admit exact algebraic compression to determinant or Pfaffian primitives~\cite{Oh2026,Valiant2002,Terhal2002}; the observables that survive noise are the observables classical estimators reproduce at matching sample complexity~\cite{Markov2025,Bosonic2025}. Together these statements close the perimeter of what NISQ hardware can deliver. The theorems do not perform the de-quantizations; concrete numerical implementations do. What the theorems supply is the reason the implementations keep succeeding. Every circuit the hardware has run well has proved, in the regimes explored to date, to sit inside that perimeter, and every observable it has estimated within shot noise has proved classically reproducible, with the single exception examined in Section~4.

The recent IBM demonstrations on the one-dimensional Heisenberg chain illustrate the loop in its purest form~\cite{Lee2026a,Lee2026b}. The Hamiltonian is integrable; Bethe solved it exactly in 1931. The transport universality classes the experiments report have been established classically, the ballistic and diffusive classes for decades and the superdiffusive KPZ class by matrix product state simulations in 2019. The structure that allowed the experiments to run on noisy hardware, namely one dimension, integrability, low-weight local observables, and short evolution times, is the same structure that makes classical MPS and Majorana-propagation methods tractable. The materials-discovery applications invoked alongside these demonstrations are two- and three-dimensional problems; the demonstrations themselves are one-dimensional by hardware necessity, which is to say the demonstrations are one or two dimensions short of the application they motivate.

\section{NISQ as historical retreat}

The threshold theorems of 1996 established that, conditional on noise satisfying specific mathematical assumptions (independence, locality, rate below threshold), arbitrarily long quantum computation is possible. The theorems are correct, of course. Whether any actual hardware satisfies their assumptions is a separate question that the theorems do not, and cannot, settle. The rub is that verifying, at scale, the correlation structure the theorems assume would itself need a quantum computer of the kind the theorems promise. The circularity has been known since the late 1990s, and it explains both the slow pace of the engineering project and the impatience of the press rooms tasked with selling its progress.

The 2018 articulation of NISQ~\cite{Preskill2018} was an acknowledgement that the conditions for the threshold theorem had not been met and would not be met soon, accompanied by a proposal that the intervening hardware be put to use anyway. The proposal had a specific technical content: that carefully chosen short circuits on noisy machines would, for some problems, outperform classical methods. The proposal was a bet on the empirical relationship between two regions of circuit-space: the region the hardware could run, and the region classical methods could not reach. The closed-loop result is that, in every regime explored to date, these two regions have turned out to be the same region.

Mele \emph{et al.}\ closes the variational programme. Oh \emph{et al.}\ reaches into the structured-input programme that motivated the trapped-ion fermionic experiments, reproducing its tractable sectors while leaving the deep interacting regime open. The bosonic and linear-optics results constrain the photonic programme. The Markovianity and geometric-locality results cover the residual cases. The constraint that defines NISQ-compatibility (logarithmic effective depth or strong algebraic structure) is the constraint the simulability results exploit. The regions of circuit-space the field has actually explored have, one after another, either fallen to classical attack or been shown to sit inside a region where classical methods follow. A future demonstration could in principle escape the current results by exploiting features they do not yet cover. After eight years in which every flagship but one has been reproduced classically, or shown to rest on classically tractable structure, or closed by a simulability theorem, within eighteen months, the burden of producing such a demonstration falls to the defenders of NISQ. The one clear exception is examined below.

\section{The diagnosis}

The standard reading of the de-quantization sequence treats each result as a contingent improvement in classical methods: someone got clever, the boundary moved, the next demonstration will choose a harder problem. The reading is incomplete. The boundary moved because the demonstrations were drawn from a region the boundary could reach. A theorist examining the trapped-ion experiment in October 2025 had, by the geometry of the situation, a finite checklist of compression strategies to try. The Pfaffian compression Oh \emph{et al.}\ used was on the list because the input states were paired, and it reached as far into the experiment as the free dynamics kept that pairing exploitable. Had the input states been unpaired the experiment would not have run on the hardware at all. The hardware selected for the structure, and the structure is what the classical method followed.

This is the diagnosis. The eight years from 2018 to 2026 were a period in which a research programme repeatedly demonstrated, on hardware whose constraints forced the demonstrations into a particular region of circuit-space, that the chosen region admitted classical simulation. Each demonstration was an empirical instance of an identity that the simulability theorems have now made formal. The de-quantizations were the only outcomes the identity permitted. Quantum advantage was not on the menu.

Two results from Google's Willow require further scrutiny. The December 2024 Willow random-circuit-sampling claim illustrates the theorem-closure half of this record: the samples have not been replicated end-to-end, and the regime the experiment occupies was closed within ten months by a general theorem on the classical simulability of noisy random circuits~\cite{RCS2025}. The October 2025 Quantum Echoes experiment~\cite{Echoes2025} is the one demonstration of the period that stands neither reproduced nor closed by theorem, and it is also the one that conforms least to the flagship template.

Without re-litigating that result, let us just note here the following. The observable is an out-of-time-order correlator selected for classical hardness. The reported values pass through a global rescaling factor obtained from error mitigation, and their validity is checked against exact answers only up to 40 qubits while the advantage is claimed at 65. The scored quantity is a standardized correlation across fifty circuits, with the mean and the variance of each dataset removed from the scoring by construction. Classical heuristics described in the paper's own supplement complicate the headline. A gate-removal tensor-network method, scored by the paper's Monte-Carlo proxy at the device's achieved accuracy target, runs in minutes to hours per circuit on Frontier against the device's 2.1 hours. That method is ruled out because the systematic error of the removal can be determined only by exact simulation or extrapolation, both unavailable at 65 qubits; the 13,000$\times$ figure is the zero-removal, memory-constrained entry of the same supplementary table. The device's own 65-qubit accuracy rests on an error-model projection and a Loschmidt-echo calibration of the twenty-fold rescaling, validated at small scale. We leave it at that.

\section{Implications}

The engineering progress of the NISQ period is real and unquestioned: coherence times, two-qubit gate fidelities, qubit counts, and control infrastructure have improved substantially. The closed-loop diagnosis targets the claim that specific NISQ demonstrations have crossed into computational regimes classical methods cannot reach, and is silent about the engineering trajectory itself. The four major US-listed quantum companies (Google, IBM, Quantinuum, IonQ) have between them issued more than thirty advantage-class announcements since 2018; the comparable Chinese efforts bring the total above forty. The sequence of de-quantizations assembled here is the technical literature catching up to that volume of claims. With this distinction in place, two implications follow.

First, the question of where unassailable quantum advantage might appear is now sharper. It must appear in a region of circuit-space the simulability theorems leave uncovered. Such regions exist as a matter of principle: the results assembled here leave intact the conjectured separation between BQP and classical polynomial time, and Shor's algorithm continues to inhabit a distinct regime that no current de-quantization addresses. The regions outside the closed loop, however, require coherence budgets the present hardware cannot supply. The exit from the loop is fault tolerance. The Mele paper makes this explicit by identifying, as the escape route from its main result, circuits engineered to act as quantum refrigerators that pump entropy out of the computation. Pumping entropy out of the computation is what error correction does. The path past the loop is the path the threshold theorems described in 1996.

In a recent paper Huang, Choi, McClean and Preskill~\cite{HuangChoi2026} propose a framework to evaluate quantum advantage claims with five criteria (predictability, typicality, robustness, verifiability, usefulness). Measured against these criteria, the NISQ-era record fares poorly. Take verifiability as an example. Agreed, theoretical computer science has produced interactive protocols~\cite{Mahadev2018} in which a classical verifier can, in principle, use cryptographic tools to check a quantum computation without simulating it. So far, however, no advantage demonstration has used one, and the sampling-style demonstrations that dominate the record cannot, by the flatness of their output distributions, be verified from samples alone~\cite{NoGo}. This emphasizes once more the empirical nature of the argument this paper makes: every demonstration actually performed to date is evidence of the mismatch between the physics and its marketing.

Second, the role of demonstration in the field requires reconsideration. A demonstration that a hardware can run a circuit, and that the circuit produces an observable matching a known answer, is a demonstration of competent diagnostic engineering. The press cycle of the past several years has conflated such results with computational advantage, and the conflation has been productive of funding without being productive of the matter of fact under investigation. Distinguishing the two locates the engineering within a programme whose ultimate test is the construction of machines capable of running circuits the simulability theorems do not reach.

\section{Moving Forward}

The hardware ran what it could. Classical algorithms simulated what the hardware ran. The press releases announced the part in between. Eight years of demonstrations whose only stable contribution to the literature has been the simulability theorems written to refute them. Quantum advantage on noisy hardware is the perpetual motion machine of theoretical physics: each new prototype runs for a few months, attracts admiring write-ups, and then a theorist notices it is plugged into the wall.

In October 2025, six months before the Oh \emph{et al.}\ result, Eisert and Preskill posted a perspective titled \emph{Mind the Gaps}~\cite{EisertPreskill2025}, introducing the acronym FASQ (Fault-Tolerant Application-Scale Quantum) and identifying four hurdles separating current devices from machines capable of useful applications. Eisert is also a co-author of the Mele \emph{et al.}\ result that closed the variational programme. The field's two most influential voices on the future of quantum computation have, between them, named the period that follows NISQ. One of them also authored a theorem that ends it.

The loop is poised to repeat under federal sponsorship. In June 2026, two executive orders committed the Department of Energy to deploying ``the world's first fault-tolerant, scientifically relevant quantum computer'' by 2028, and the DOE's request for information of May 2026 sets the targets: 150 to 250 logical qubits, a universal instruction set, circuits of $10^5$ hard gates at a logical error rate of $10^{-8}$~\cite{DOE2026}.

The two adjectives in the DOE's phrase do separate work, and the coverage of the orders already displays the equivocation~\cite{Spectrum2026}. Asked about the 2028 machine, Jay Sau of Maryland cannot name a scientific question it answers absent a breakthrough in device quality; Pranav Gokhale of Infleqtion places the advantage at 100 logical qubits, on models of magnetism and high-temperature superconductivity; Edward Parker of RAND supplies the reconciling observation that the logical qubits in question will be ``somewhat more stable than their underlying constituent physical qubits, but still noisy.''

These verdicts describe the same hardware with one condition turned on or off, and a machine meeting the RFI's description in 2028 will be one of three things.
\begin{itemize}
\item The first is a fault-tolerant computer: logical qubits that actively pump entropy out of the computation, free of the conditions NISQ-era hardware results depend on, namely postselection and decoders trained on assumed noise models. These conditions being gone is the same event as the machine being done, and such a machine clears the bar the threshold theorems set in 1996.

\item The second is a NISQ device with a larger qubit count: logical qubits merely stable enough for a few tailored tasks, running under those same conditions, to which the Mele theorem applies as written; its typical circuits are logarithmically shallow and its expectation values classically estimable.

\item The third is a physics experiment: a noisy device operated as the system under study, measuring its own dynamics, as in Hamiltonian learning. This is legitimate science with its own standards and no advantage claim attached, and it already has a delivered instance. Quantum Echoes, examined above, is a measurement of the device's own scrambling dynamics, dressed as a step into computational regimes classical methods cannot reach.

\end{itemize}

One quantity decides among the three, independently of the label on the qubits: the achieved logical error rate times the circuit depth, which must stay below the erasure bound the simulability theorems set. The solicitation the RFI prefigures can enforce that arithmetic, pricing the deliverable in the RFI's own units of logical error rate and hard-gate count, or it can let ``scientifically relevant'' and ``fault-tolerant'' ride together. In the latter case a machine of the second or third kind will be scored as the first in 2028, the funding will renew, and the record assembled in this paper will acquire a second decade.

\end{document}